\documentclass[aps,pra,reprint,superscriptaddress,nofootinbib]{revtex4-1}

\usepackage[utf8]{inputenc}
\usepackage[T1]{fontenc}
\usepackage{lmodern}
\usepackage{microtype}

\usepackage{amsmath}
\usepackage{graphicx}
\usepackage{hyperref}
\hypersetup{
  colorlinks=true,
  citecolor=blue,
  linkcolor=blue,
  urlcolor=blue
}

\usepackage{comment}
\usepackage[braket,qm]{qcircuit}

\DeclareMathOperator*{\argmin}{arg\,min}

\begin{document}

\title{Unfolding Quantum Computer Readout Noise}

\author{Benjamin Nachman}
\email{bpnachman@lbl.gov}
\affiliation{Physics Division, Lawrence Berkeley National Laboratory, Berkeley, CA 94720, USA}

\author{Miroslav Urbanek}
\affiliation{Computational Research Division, Lawrence Berkeley National Laboratory, Berkeley, CA 94720, USA}

\author{Wibe A. de Jong}
\affiliation{Computational Research Division, Lawrence Berkeley National Laboratory, Berkeley, CA 94720, USA}

\author{Christian W. Bauer}
\affiliation{Physics Division, Lawrence Berkeley National Laboratory, Berkeley, CA 94720, USA}

\begin{abstract}
In the current era of noisy intermediate-scale quantum (NISQ) computers, noisy qubits can result in biased results for early quantum algorithm applications.  This is a significant challenge for interpreting results from quantum computer simulations for quantum chemistry, nuclear physics, high energy physics, and other emerging scientific applications.  An important class of qubit errors are readout errors.  The most basic method to correct readout errors is matrix inversion, using a response matrix built from simple operations to probe the rate of transitions from known initial quantum states to readout outcomes.  One challenge with inverting matrices with large off-diagonal components is that the results are sensitive to statistical fluctuations.  This challenge is familiar to high energy physics, where prior-independent regularized matrix inversion techniques (`unfolding') have been developed for years to correct for acceptance and detector effects when performing differential cross section measurements.  We study various unfolding methods in the context of universal gate-based quantum computers with the goal of connecting the fields of quantum information science and high energy physics and providing a reference for future work.  The method known as iterative Bayesian unfolding is shown to avoid pathologies from commonly used matrix inversion and least squares methods.
\end{abstract}

\date{\today}
\maketitle

\section{Introduction}

While quantum algorithms are promising techniques for a variety of scientific and industrial applications, current challenges limit their immediate applicability.  One significant limitation is the rate of errors and decoherence in noisy intermediate-scale quantum (NISQ) computers~\cite{Preskill2018quantumcomputingin}. Mitigating errors is hard in general because quantum bits (`qubits') cannot be copied~\cite{Park1970,Wootters:1982zz,DIEKS1982271}. An important family of errors are readout errors. They typically arise from two sources: (1) measurement times are significant in comparison to decoherence times and thus a qubit in the $\ket{1}$ state can decay to the $\ket{0}$ state during a measurement, and (2) probability distributions of measured physical quantities that correspond to the $\ket{0}$ and $\ket{1}$ states have overlapping support and there is a small probability of measuring the opposite value. The goal of this paper is to investigate methods for correcting these readout errors.  This is complementary to efforts for gate error corrections.  One strategy for mitigating such errors is to build in error correcting components into quantum circuits.  Quantum error correction~\cite{0904.2557,Devitt_2013,RevModPhys.87.307,2013qec..book.....L,Nielsen:2011:QCQ:1972505} is a significant challenge because qubits cannot be cloned~\cite{Park1970,Wootters:1982zz,DIEKS1982271}.  This generates a signficant overhead in the additional number of qubits and gates required to detect or correct errors.  Partial error detection/correction has been demonstrated for simple quantum circuits~\cite{errorcorrecting,PhysRevA.97.052313,Barends2014,Kelly2015,Linke:2017, Takita:2017, Roffe:2018, Vuillot:2018, Willsch:2018,Harper:2019}, but complete error correction is infeasible for current qubit counts and moderately deep circuits.  As a result, many studies with NISQ devices use the alternative zero noise extrapolation strategy whereby circuit noise is systematically increased and then extrapolated to zero~\cite{Kandala:2019,PhysRevX.7.021050,PhysRevLett.119.180509,PhysRevX.8.031027,Dumitrescu:2018,PhysRevLett.119.180509,RIIM}.  Ultimately, both gate and readout errors must be corrected for a complete measurement and a combination of strategies may be required. 

%One approach that is currently used for correcting readout noise is to invert the confusion matrix, which codifies the probability for a qubit in one state to be measured in another state.  Should cite 1904.10440 (see appendix) and https://github.com/Qiskit/qiskit-ignis.

%Paragraph about measurements in high energy physics.

Correcting measured histograms for the effects of a detector has a rich history in image processing, astronomy, high energy physics (HEP), and beyond.  In the latter, the histograms represent binned differential cross sections and the correction is called \textit{unfolding}.  Many unfolding algorithms have been proposed and are currently in use by experimental high energy physicists (see, e.g., Ref.~\cite{Cowan:2002in,Blobel:2203257,doi:10.1002/9783527653416.ch6} for reviews).  One of the goals of this paper is to introduce these methods and connect them with current algorithms used by the quantum information science community.

\begin{figure}
\centering
\includegraphics[width=\linewidth]{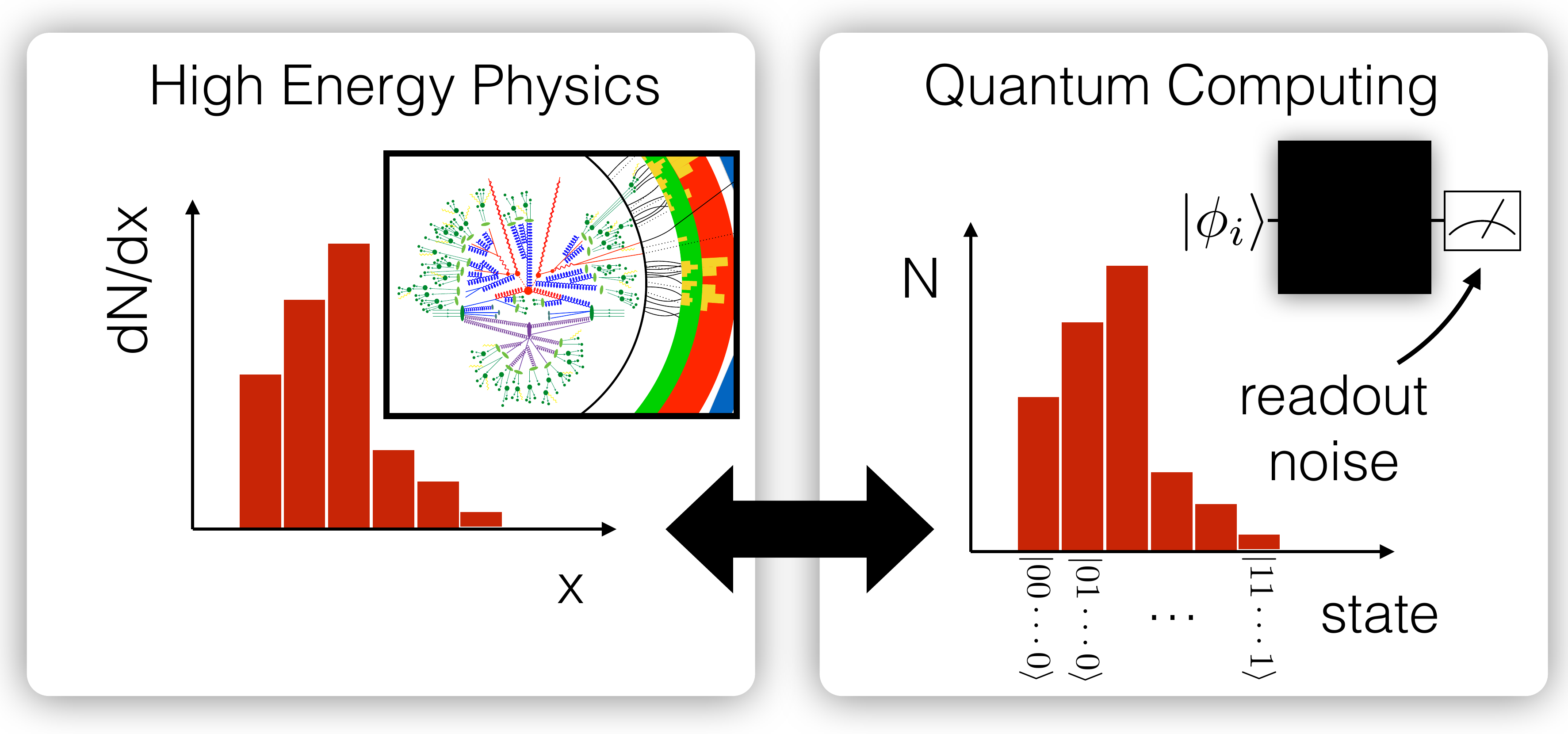}
\caption{A schematic diagram illustrating the connection between binned differential cross section measurements in high energy physics (left) and interpreting the output of repeated measurements from quantum computers (right).}
\label{lab:schematic}
\end{figure}

Quantum readout error correction can be represented as histogram unfolding where each bin corresponds to one of the possible $2^{n_\text{qubit}}$ configurations where $n_\text{qubit}$ is the number of qubits (Fig.~\ref{lab:schematic}).  Correcting readout noise is a classical problem (though there has been a proposal to do it with quantum annealing~\cite{Cormier:2019kcq}), but relies on calibrations or simulations from quantum hardware.  Even though discussions of readout errors appear in many papers (see, e.g., Ref.~\cite{Kandala:2017,Dumitrescu:2018,Klco:2019xro,YeterAydeniz:2019}), we are not aware of any dedicated study of unfolding methods for quantum information science (QIS) applications.  Furthermore, current QIS methods have pathologies that can be avoided with techniques from HEP.

This paper is organized as follows.  Sec.~\ref{sec:intro} introduces the unfolding setup. Various techniques are described in Sec.~\ref{sec:unfolding}.  The core input for unfolding is the matrix of probabilities relating the true and measured spectra that is discussed in Sec.~\ref{sec:responsematrix}. Representative examples are presented in Sec.~\ref{sec:results} and a description of sources of uncertainty in Sec.~\ref{sec:uncerts}. The discussion in Sec.~\ref{sec:discussion} is followed by a summary and outlook in Sec.~\ref{sec:conclusions}.

\section{The unfolding challenge}
\label{sec:intro}

Let $t$ be a vector that represents the true bin counts before the distortions from detector effects (HEP) or readout noise (QIS).  The corresponding measured bin counts are denoted by $m$.  These vectors are related by a \textit{response matrix} $R$ as $m=Rt$ where $R_{ij}=\Pr(m=i|t=j)$.  In HEP, the matrix $R$ is usually estimated from detailed detector simulations while in QIS, $R$ is constructed from measurements of computational basis states.  The response matrix construction is discussed in Sec.~\ref{sec:responsematrix}.

\begin{figure}
\centering
\includegraphics[width=\linewidth]{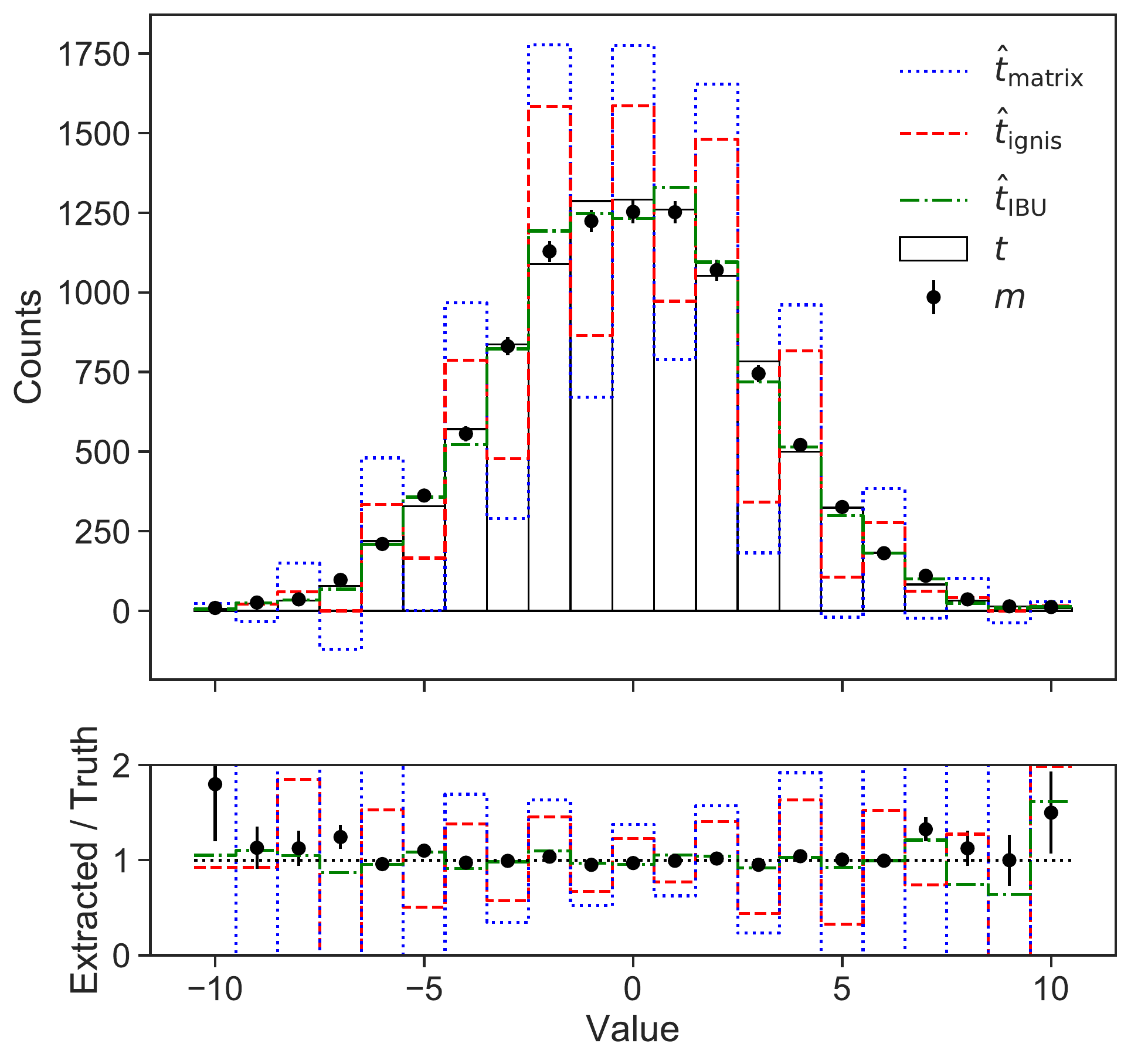}
\caption{A comparison of unfolding techniques for a Gaussian example and the $R$ matrix from Eq.~\eqref{eq:pathologicalexample}.  The symbols $t$ and $m$ denote the true and measured probability mass functions, respectively.   Simple matrix inversion is represented by $\hat{t}_\text{matrix}$ (see Sec.~\ref{sec:intro}).  The \texttt{ignis} and IBU methods are represented by $\hat{t}_\text{ignis}$ and $\hat{t}_\text{IBU}$, respectively (see Sec.~\ref{sec:unfolding}).  For this example, $||m||_1=10^4$ and $R$ is assumed to be known exactly.  The IBU method uses 10 iterations and a uniform prior (other iterations choices are studied in App.~\ref{sec:iterations}).  The simulation used in this plot is based on standard Python functions and does not use a quantum computer simulator (see instead Fig.~\ref{lab:pathologicalexampleqiskit}).}
\label{lab:matrixinversion}
\end{figure}

The most naive unfolding procedure would be to simply invert the matrix $R$: $\hat{t}_\text{matrix}=R^{-1}m$.  However, simple matrix inversion has many known issues.  Two main problems are that $\hat{t}_\text{matrix}$ can have unphysical entries and that statistical uncertainties in $R$ can be amplified and can result in oscillatory behavior.  For example, consider the case
\begin{align}
R=\begin{pmatrix}1-\epsilon & \epsilon \\ \epsilon & 1-\epsilon\end{pmatrix},
\end{align}
where $0<\epsilon <1/2$.  Then, $\text{Var}(\hat{t}_\text{matrix})\propto 1/\det(R)=1/(1-2\epsilon)\rightarrow \infty$ as $\epsilon\rightarrow 1/2$.  As a generalization of this example to more bins (from Ref.~\cite{Blobel:157405}), consider a response matrix with a symmetric probability of migrating one bin up or down,
\begin{align}
\label{eq:pathologicalexample}
R=\begin{pmatrix}1-\epsilon & \epsilon & 0 & \cdots \\ \epsilon & 1-2\epsilon & \epsilon & \cdots \\ 0 &  \epsilon & 1-2\epsilon & \cdots \\ \vdots & \vdots & \vdots & \ddots \end{pmatrix}.
\end{align}
Unfolding with the above response matrix and $\epsilon=25\%$ is presented in Fig.~\ref{lab:matrixinversion}.  The true bin counts have the Gaussian distribution with a mean of zero and a standard deviation of 3.  The values are discretized with 21 uniform unit width bins spanning the interval $[-10,10]$.  The leftmost bin correspond to the first index in $m$ and $t$. The indices are monotonically incremented with increasing bin number.  The first and last bins include underflow and overflow values, respectively.  Due to the symmetry of the migrations, the true and measured distributions are nearly the same.  For this reason, the measured spectrum happens to align with the true distribution; the optimal unfolding procedure should be the identity.  The significant off-diagonal components result in an oscillatory behavior and the statistical uncertainties are also amplified by the limited size of the simulation dataset used to derive the response matrix.

\section{Unfolding methods}
\label{sec:unfolding}

The fact that matrix inversion can result in unphysical outcomes ($\hat{t}_i<0$ or $\hat{t}_i>||t||_1$, where $||x||_1=\sum_i|x_i|$ is the $L^1$ norm) is often unacceptable.  One solution is to find a vector that is as close to $\hat{t}$ as possible, but with physical entries.  This is the method implemented in \texttt{qiskit-ignis}~\cite{qiskit,ignis}, a widely used quantum computation software package.  The \texttt{ignis} method\footnote{This has been recently studied by other authors as well, see e.g. Ref.~\cite{1904.11935,1907.08518}.} solves the optimization problem
\begin{align}
\label{eq:ignis}
\hat{t}_\text{ignis}=\argmin\limits_{t':\,||t'||_1=||m||_1,\,t_i' > 0}||m-Rt'||^2.
\end{align}
Note that the norm invariance is also satisfied for simple matrix inversion: $||\hat{t}_\text{matrix}||_1=||m||_1$ by construction because $||Rx||_1=||x||_1$ for all $x$ so in particular for $y=R^{-1}m$, $||Ry||_1=||y||_1$ implies that $||m||_1=||R^{-1}m||_1$.  This means that $\hat{t}_\text{ignis}=\hat{t}_\text{matrix}$ whenever the latter is non-negative and so $\hat{t}_\text{ignis}$ inherits some of the pathologies of $\hat{t}_\text{matrix}$.

Three commonly used unfolding methods in HEP\footnote{There are other less widely used methods such as fully Bayesian unfolding~\cite{Choudalakis:2012hz} and others~\cite{Gagunashvili:2010zw,Glazov:2017vni,Datta:2018mwd,Aslan:2003vu,Lindemann:1995ut}.} are Iterative Bayesian\footnote{Even though this method call for the repeated application of Bayes' theorem, this is not a Bayesian method as there is no prior/posterior over possible distributions, only a prior for initialization.} unfolding (IBU)~\cite{DAgostini:1994fjx} (also known as Richardson-Lucy deconvolution~\cite{1974AJ.....79..745L,Richardson:72}), Singular Value Decomposition (SVD) unfolding~\cite{Hocker:1995kb}, and \texttt{TUnfold}~\cite{Schmitt:2012kp}.  \texttt{TUnfold} is similar to Eq.~\eqref{eq:ignis}, but imposes further regularization requirements in order to avoid pathologies from matrix inversion.  The SVD approach applies some regularization directly on $R$ before applying matrix inversion.  The focus of this paper will be on the widely used IBU method, which avoids fitting\footnote{Reference~\cite{DAgostini:1994fjx} also suggested that one could combine the iterative method with smoothing from a fit in order to suppress amplified statistical fluctuations.} and matrix inversion altogether with an iterative approach.

%RooUnfold~\cite{Adye:2011gm}. Bayesian unfolding~\cite{DAgostini:1994fjx} (also known as Richardson-Lucy deconvolution~\cite{1974AJ.....79..745L,Richardson:72}), SVD unfolding~\cite{Hocker:1995kb}, TUnfold~\cite{Schmitt:2012kp}, qiskit-ignis~\cite{ignis}.

Given a prior truth spectrum $t_i^0 = \Pr(\text{truth is $i$})$, the IBU technique proceeds according to the equation
\begin{align}
\label{eq:unfolding}
\begin{split}
t_i^{n+1} &= \sum_j \Pr(\text{truth is $i$} | \text{measure $j$}) \times m_j \\
&= \sum_j \frac{R_{ji} t_i^n}{\sum_k R_{jk} t_k^n} \times m_j,
\end{split}
\end{align}
where $n$ is the iteration number and one iterates a total of $N$ times.  The advantage of Eq.~\eqref{eq:unfolding} over simple matrix inversion is that the result is a probability (non-negative and unit measure) when\footnote{In HEP applications, $m$ can have negative entries resulting from background subtraction.  In this case, the unfolded result can also have negative entries.} $m\geq 0$.  The parameters $t_i^0$ and $N$ must be specified ahead of time.  A common choice for $t_i^0$ is the uniform distribution.  The number of iterations needed to converge depends on the desired precision, how close $t_i^0$ is to the final distribution, and the importance of off-diagonal components in $R$.  In practice, it may be desirable to choose a relatively small $N$ prior to convergence to regularize the result.  Typically, $\lesssim\mathcal{O}(10)$ iterations are needed.

In addition to the $\hat{t}_\text{matrix}$ and $\hat{t}_\text{ignis}$ approaches discussed in the previous section, Fig.~\ref{lab:matrixinversion} shows the IBU result with $N=10$.  Unlike the $\hat{t}_\text{matrix}$ and $\hat{t}_\text{ignis}$ results, the $\hat{t}_\text{IBU}$ does not suffer from rapid oscillations and like $\hat{t}_\text{ignis}$, is non-negative.  Analogous results for quantum computer simulations will be presented in Sec.~\ref{sec:results}.

%We construct $R_{ij}$ by preparing $2^m$ calibration circuits where each qubit configuration is constructed with $X$ gates. The entries of $R_{ij}$ are the fraction of measurements that qubit configuration $i$ is observed in configuration $j$. We use a uniform distribution as the initial spectrum $t_i^0$. 

\section{Constructing the response matrix}
\label{sec:responsematrix}

\begin{figure}
\[
\Qcircuit @C=0.5em @R=0.8em @!R{
\lstick{\ket{0}} & \qw & \qw & \qw & \meter &&&&&&& \lstick{\ket{0}} & \qw & \gate{X} & \qw & \meter &&&& \cdots &&&&&&&\lstick{\ket{0}} & \qw & \gate{X} & \qw & \meter \\
\lstick{\ket{0}} & \qw & \qw & \qw & \meter &&&&&&& \lstick{\ket{0}} & \qw & \qw & \qw & \meter &&&& \cdots &&&&&&&\lstick{\ket{0}} & \qw & \gate{X} & \qw & \meter \\
\lstick\vdots & & \vdots & & \vdots &&&&&&& \lstick\vdots & & \vdots & & \vdots &&&&&&&&&&& \lstick\vdots & & \vdots & & \vdots \\
\lstick{\ket{0}} & \qw & \qw & \qw & \meter &&&&&&& \lstick{\ket{0}} & \qw & \qw& \qw & \meter &&&& \cdots &&&&&&&\lstick{\ket{0}} & \qw & \gate{X} & \qw & \meter \\
}
\]

\caption{The set of $2^{n_\text{qubit}}$ calibration circuits.}
\label{fig:calibrationcircuits}
\end{figure}
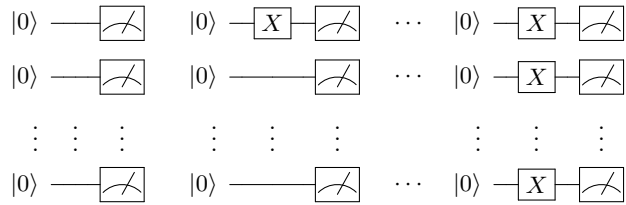

In practice, the $R$ matrix is not known exactly and must be measured for each quantum computer and set of operating conditions.  One way to measure $R$ is to construct a set of $2^{n_\text{qubit}}$ \textit{calibration circuits} as shown in Fig.~\ref{fig:calibrationcircuits}.  Simple $X$ gates are used to prepare all of the possible qubit configurations and then they are immediately measured.

Fig.~\ref{lab:responsetokyo} shows the $R$ matrix from 5 qubits of the IBM Q Johannesburg machine (see Appendix~\ref{sec:alternativeconfigurations} for details).  As expected, there is a significant diagonal component that corresponds to cases when the measured and true states are the same.  However, there are significant off-diagonal components, which are larger towards the right when more configurations start in the one state.  The diagonal stripes with transition probabilities of about $5-7\%$ are the result of the same qubit flipping from $0\leftrightarrow 1$.  This matrix is hardware-dependent and its elements can change over time due to calibration drift.  Machines with higher connectivity have been observed to have more readout noise.

\begin{figure}
\centering
\includegraphics[width=\linewidth]{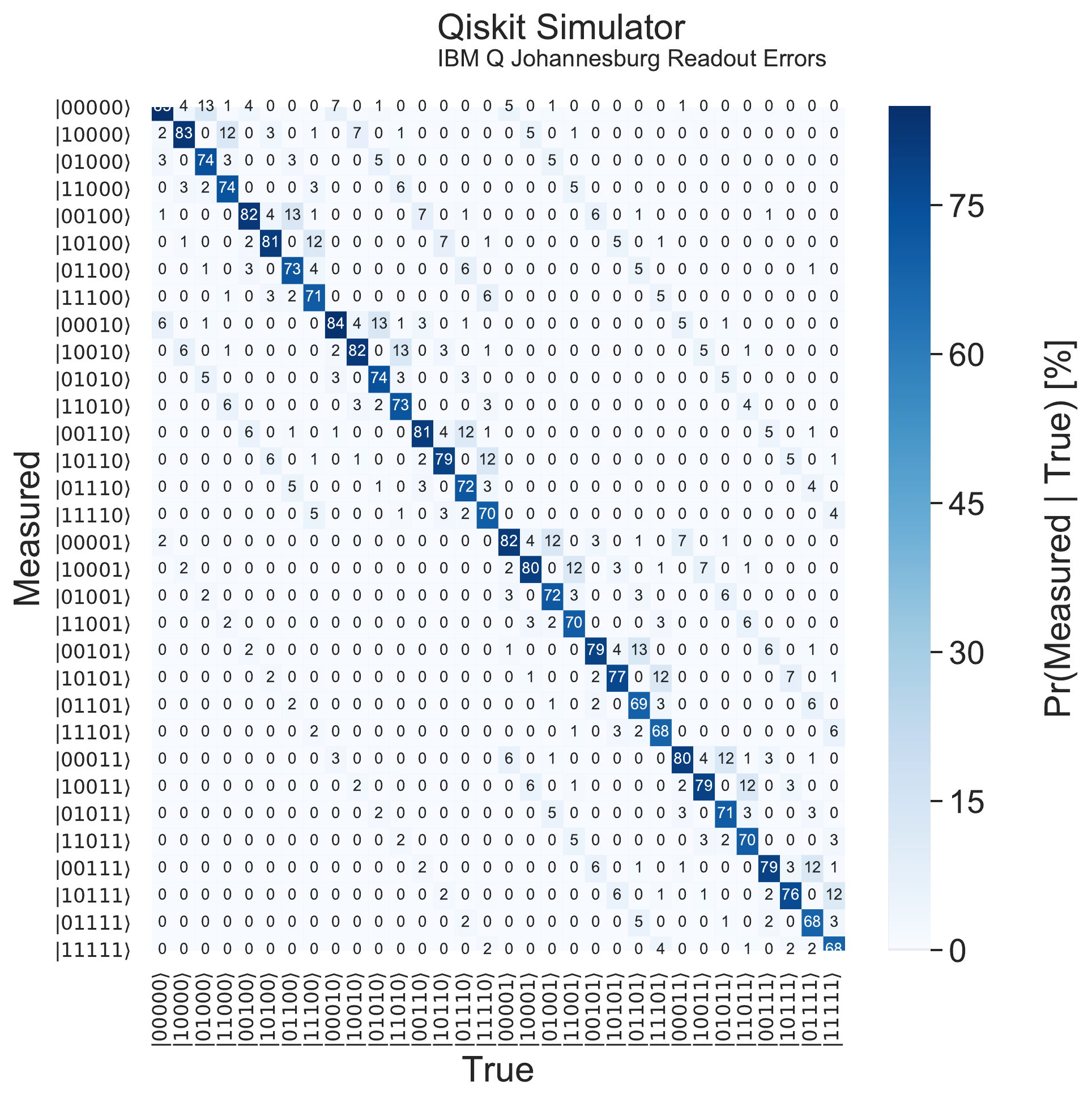} %ibmqmat.pdf}
\caption{An example $R$ matrix from 5 qubits of the IBM Q Johannesburg machine using 8192 shots for each of the $2^5$ possible states.  Note that this matrix depends on the calibration quality, so while it is representative, it is not precisely the version valid for all measurements made on this hardware.}
\label{lab:responsetokyo}
\end{figure}

\begin{figure}
\centering
\includegraphics[width=\linewidth]{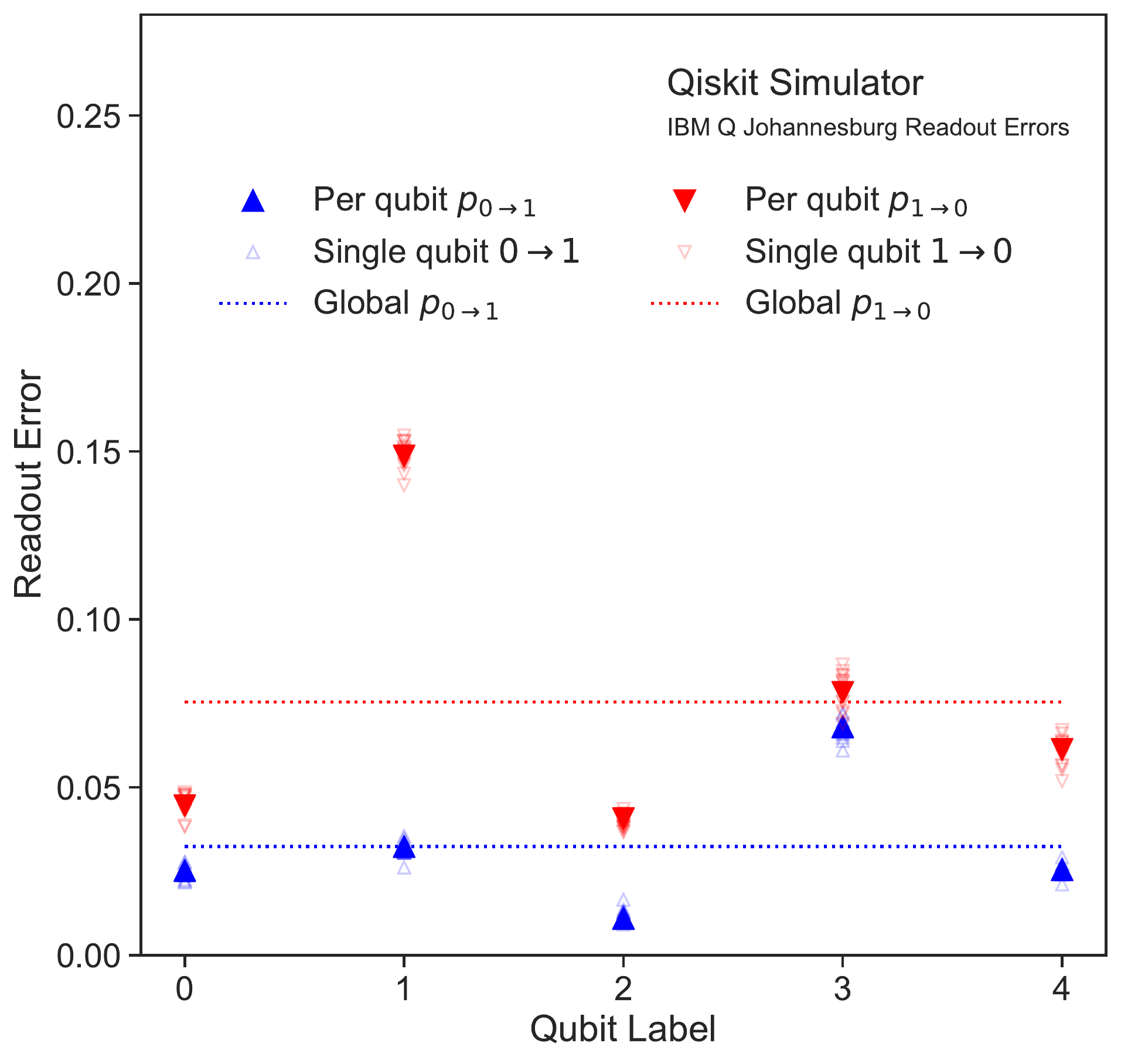} %ibmqtest.pdf}
\caption{Test of universality in the response matrix $R$.  Horizontal dotted lines show the result of a global fit to $p_{0\rightarrow1}$ and $p_{1\rightarrow 0}$ assuming an equal rate of readout errors for each qubit. Big filled markers represent a fit with independent values for each qubit.  Small semi-transparent open markers show the $2^{n_\text{qubit}-1}$ transition probabilities for each qubit when the other $n_\text{qubit}-1$ qubits are unchanged between truth and measured states.}
\label{lab:universality}
\end{figure}

Fig.~\ref{lab:universality} explores the universality of qubit migrations across $R$.  If every qubit was identical and there were no effects from the orientation and connectivity of the computer, then one may expect that $R$ can actually be described by just two numbers $p_{0\rightarrow1}$ and $p_{1\rightarrow 0}$, the probability for a zero to be measured as a one and vice versa.  These two numbers are extracted from $R$ by performing the fit in Eq.~\ref{eq:fit}:

\begin{align}
\label{eq:fit}
\min_{\substack{p_{0\rightarrow1} \\ p_{1\rightarrow 0}}}\sum_{ij}|R_{ij}-p_{0\rightarrow1}^\alpha(1-p_{0\rightarrow1})^{\alpha'} p_{1\rightarrow 0}^\beta(1-p_{1\rightarrow 0})^{\beta'}|^2,
\end{align}

\noindent where $\alpha$ is the number of qubits corresponding to the response matrix entry $R_{ij}$ that flipped from $0$ to $1$, $\alpha'$ is the number that remained as a $0$ and $\beta, \beta'$ are the corresponding entries for a one state; $\alpha+\alpha'+\beta+\beta'=n_\text{qubit}$.  For example, if the $R_{ij}$ entry corresponds to the state $\ket{01101}$ migrating to $\ket{01010}$, then $\alpha=1$, $\alpha'=1$, $\beta=2$, and $\beta'=1$.

A global fit to these parameters for the Johannesburg machine results in $p_{0\rightarrow1}\approx 3.2\%$ and $p_{1\rightarrow 0}\approx 7.5\%$.  In reality, the qubits are not identical and so one may expect that $p_{0\rightarrow1}$ and $p_{1\rightarrow0}$ depend on the qubit.  A fit to $n_\text{qubit}$ values for $p_{0\rightarrow1}$ and $p_{1\rightarrow0}$ (Eq.~\eqref{eq:fit}, but fitting 10 parameters instead of 2) are shown as filled triangles in Fig.~\ref{lab:universality}.  While these values cluster around the universal values (dotted lines), the spread is a relative 50\% for $p_{0\rightarrow1}$ and 60\% for $p_{1\rightarrow0}$.  Furthermore, the transition probabilities can depend on the values of neighboring qubits.  The open markers in Fig.~\ref{lab:universality} show the transition probabilities for each qubit with the other $n_\text{qubit}-1$ qubits held in a fixed state. In other words, the $2^{n_\text{qubit}-1}$ open triangles for each qubit show
\begin{align}
\Pr(\ket{q_0,\dots,q_i,\dots,q_{n_\text{qubit}}}\rightarrow\ket{q_0,\dots,q_i',\dots,q_{n_\text{qubit}}})
\,,
\end{align}
where $(q_i,q_i')\in\{(0,1),(1,0)\}$ with the other qubits $q_j$ held in a fixed configuration.  The spread in these values\footnote{This spread has a contribution from connection effects, but also from Poisson fluctuations as each measurement is statistically independent.} is smaller than the variation in the solid markers, which indicates that per-qubit readout errors are likely sufficient to capture most of the salient features of the response matrix.  However, this is hardware dependent and higher connectivity computers may depend more on the state of neighboring qubits.

%the connectivity with neighboring qubits plays an important role in the readout error.

Constructing the entire response matrix requires exponential resources in the number of qubits.  While the measurement of $R$ only needs to be performed once per quantum computer per operational condition, this can be untenable when $n_\text{qubit}\gg 1$.  The tests above indicate that a significant fraction of the $2^{n_\text{qubit}}$ calibration circuits may be required for a precise measurement of $R$.  Sub-exponential approaches may be possible and will be studied in future work.

\section{Representative Results}
\label{sec:results}

For the results presented here, we simulate a quantum computer using \texttt{qiskit-terra} 0.9.0, \texttt{qiskit-aer} 0.2.3, and \texttt{qiskit-ignis} 0.1.1~\cite{qiskit}\footnote{Multi-qubit readout errors are not supported in all versions of \texttt{qiskit-aer} and \texttt{qiskit-ignis}. This work uses a custom measure function to implement the response matrices.}.  We choose a Gaussian distribution as the true distribution, as this is ubiquitous in quantum mechanics as the ground state of the harmonic oscillator\footnote{An alternative distribution that is mostly zero with a small number of spikes is presented in Appendix~\ref{sec:alternative}.}.  This system has been recently studied in the context quantum field theory as a benchmark $0+1$ dimensional non-interacting scalar field theory~\cite{Jordan:2017lea,Jordan:2011ci,Jordan:2011ne,Jordan:2014tma,10.5555/3179430.3179434,PhysRevLett.121.110504,Macridin:2018oli,Klco:2018zqz}. In practice, all of the qubits of a system would be entangled to achieve the harmonic oscillator wave function.  However, this is unnecessary for studying readout errors, which act at the ensemble level.  The Gaussian is mapped to qubits using the following map:
\begin{align}
t(b)\propto\exp\left[-\frac{1}{2\sigma }\left(b-2^{n_\text{qubit}-1}\right)^2\right],
\end{align}
where $b$ is the binary representation of a computational basis state, i.e., $\ket{00000}\mapsto 0$ and $\ket{00011}\mapsto 3$.  For the results shown below, $\sigma=3.5$.

\begin{figure}
\centering
\includegraphics[width=\linewidth]{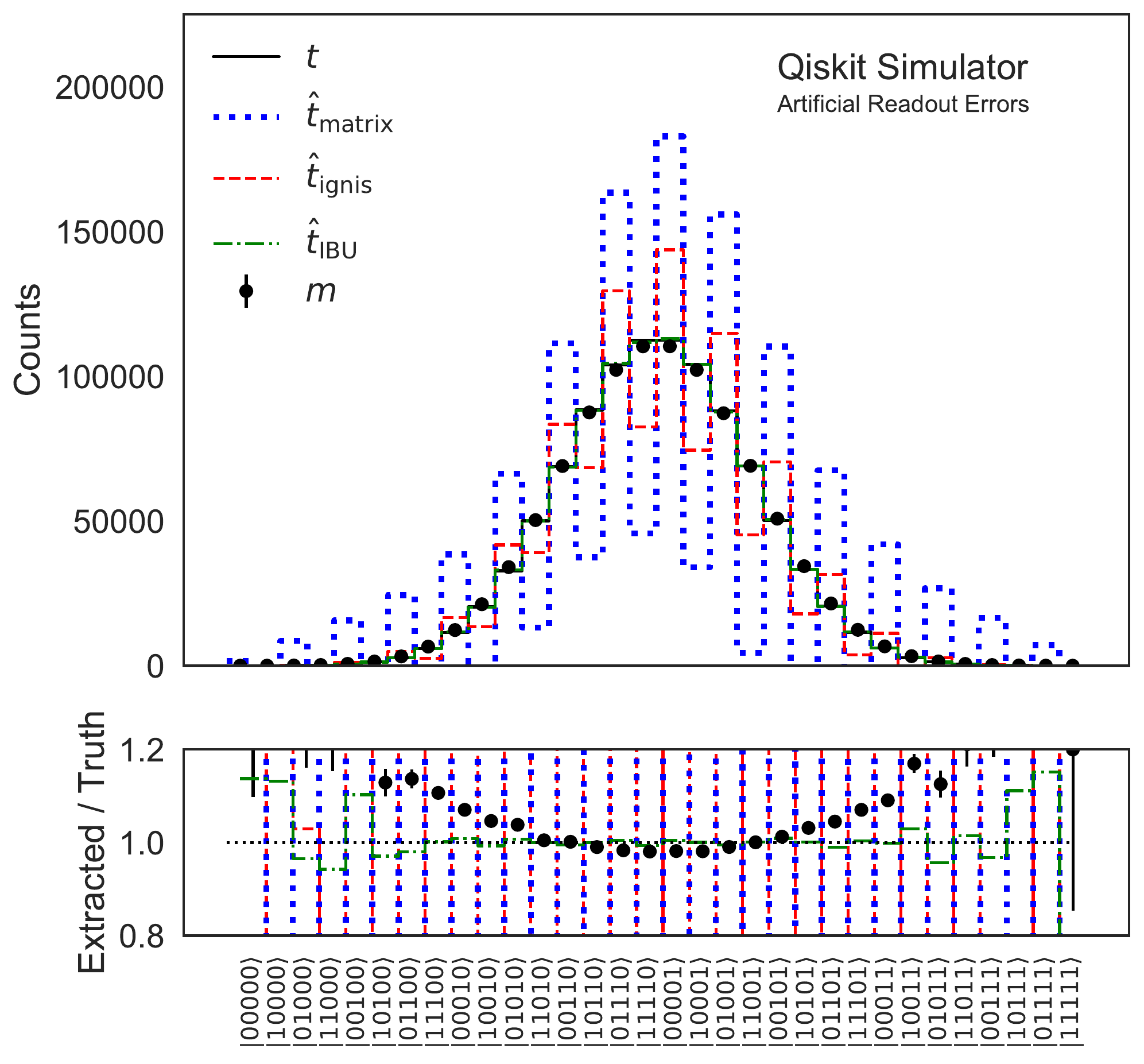}
\caption{The measurement of a Gaussian distribution (ground state of harmonic oscillator) using the pathological response matrix from Sec.~\ref{sec:unfolding}.  One million total shots are used both to sample from $m$ and to construct $R$.  The IBU method uses 10 iterations.}
\label{lab:pathologicalexampleqiskit}
\end{figure}

As a first test, the pathological response matrix introduced in Sec.~\ref{sec:unfolding} is used to illustrate a failure mode of the matrix inversion and \texttt{ignis} approaches.  Fig.~\ref{lab:pathologicalexampleqiskit} compares $\hat{t}_\text{matrix}$, $\hat{t}_\text{ignis}$, and $\hat{t}_\text{IBU}$ for a 4-qubit Gaussian distribution.  As the matrix inversion result is already non-negative, the $\hat{t}_\text{ignis}$ result is nearly identical to $\hat{t}_\text{matrix}$.  Both of these approaches show large oscillations.  In contrast, the IBU method with 10 iterations is nearly the same as the truth distribution.  This result is largely insensitive to the choice of iteration number, though it approaches the matrix inversion result for more than about a thousand iterations.  This is because the IBU method converges to a maximum likelihood estimate~\cite{4307558}, which in this case aligns with the matrix inversion result.  If the matrix inversion would have negative entries, then it would differ from the asymptotic limit of the IBU method, which is always non-negative.

\begin{figure}
\centering
\includegraphics[width=0.975\linewidth]{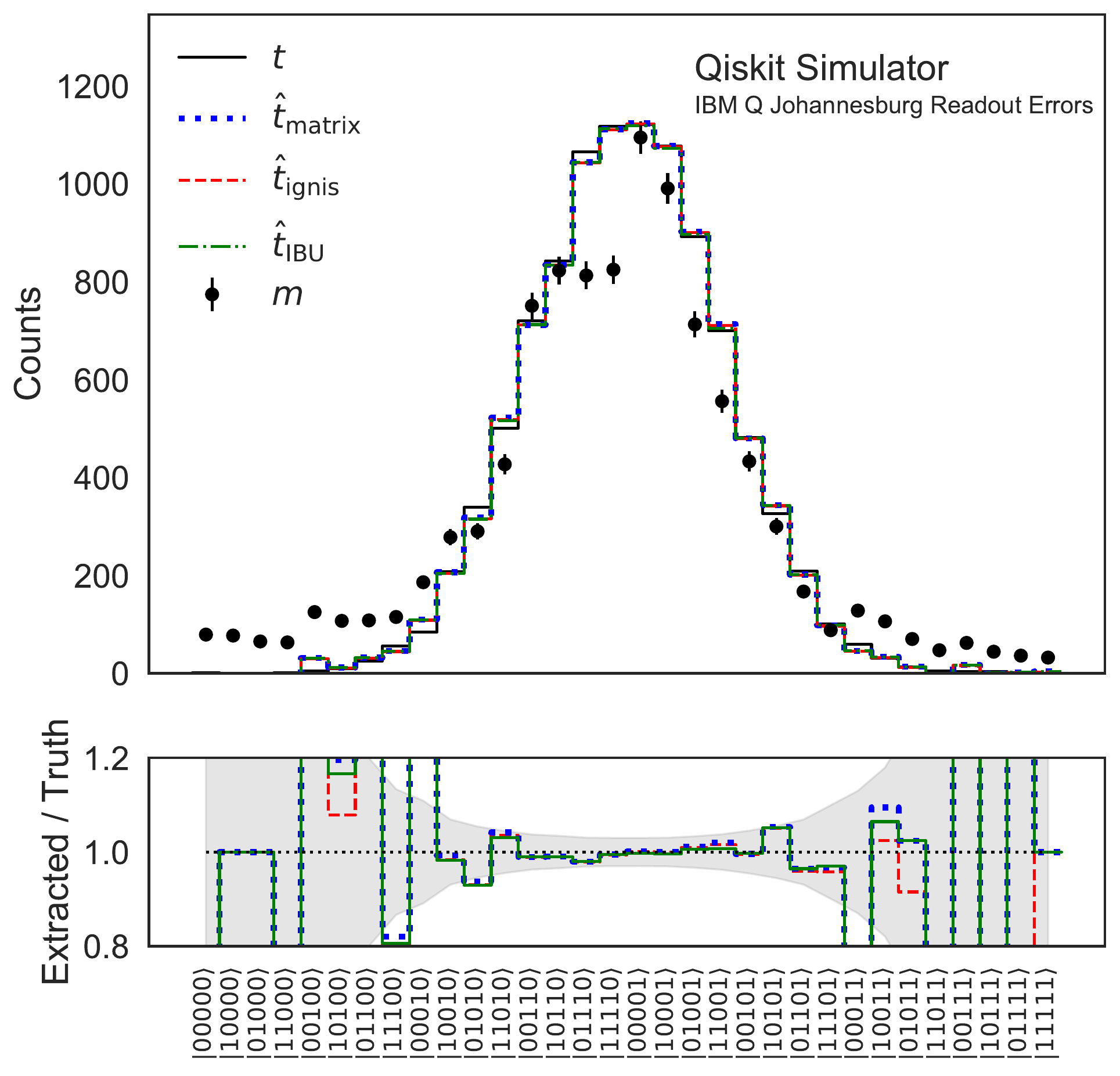}
\caption{The measurement of a Gaussian distribution using the response matrix from the IBM Q Johannesburg machine described in Sec.~\ref{sec:responsematrix}.  One million total shots are used construct $R$ and $10^4$ are used for $t$ and $m$.   The IBU method uses 100 iterations.   The significant deviations on the far left and right of the distributions are due in part to large statistical fluctuations, where the counts are low.  The uncertainty band in the ratio is the statistical uncertainty on $t$.}
\label{lab:nonpathologicalexampleqiskit}
\end{figure}

While Fig.~\ref{lab:pathologicalexampleqiskit} is illustrative for exposing a potential failure mode of matrix inversion and the \texttt{ignis} method, it is also useful to consider a realistic response matrix.  Fig.~\ref{lab:nonpathologicalexampleqiskit} uses the same Gaussian example as above, but for the IBM Q Johannesburg response matrix described in Sec.~\ref{sec:responsematrix}.  Even though the migrations are large, the pattern is such that all three methods qualitatively reproduce the truth distribution.  The dip in the middle of the distribution is due to the mapping between the Gaussian and qubit states: all of the state to the left of the center have a zero as the last qubit while the ones on the right have a one as the last qubit.  The asymmetry of $0\rightarrow 1$ and $1\rightarrow 0$ induces the asymmetry in the measured spectrum.  For example, it is much more likely to migrate from any state to $\ket{00000}$ than to $\ket{11111}$.

\begin{figure}
\centering
\includegraphics[width=\linewidth]{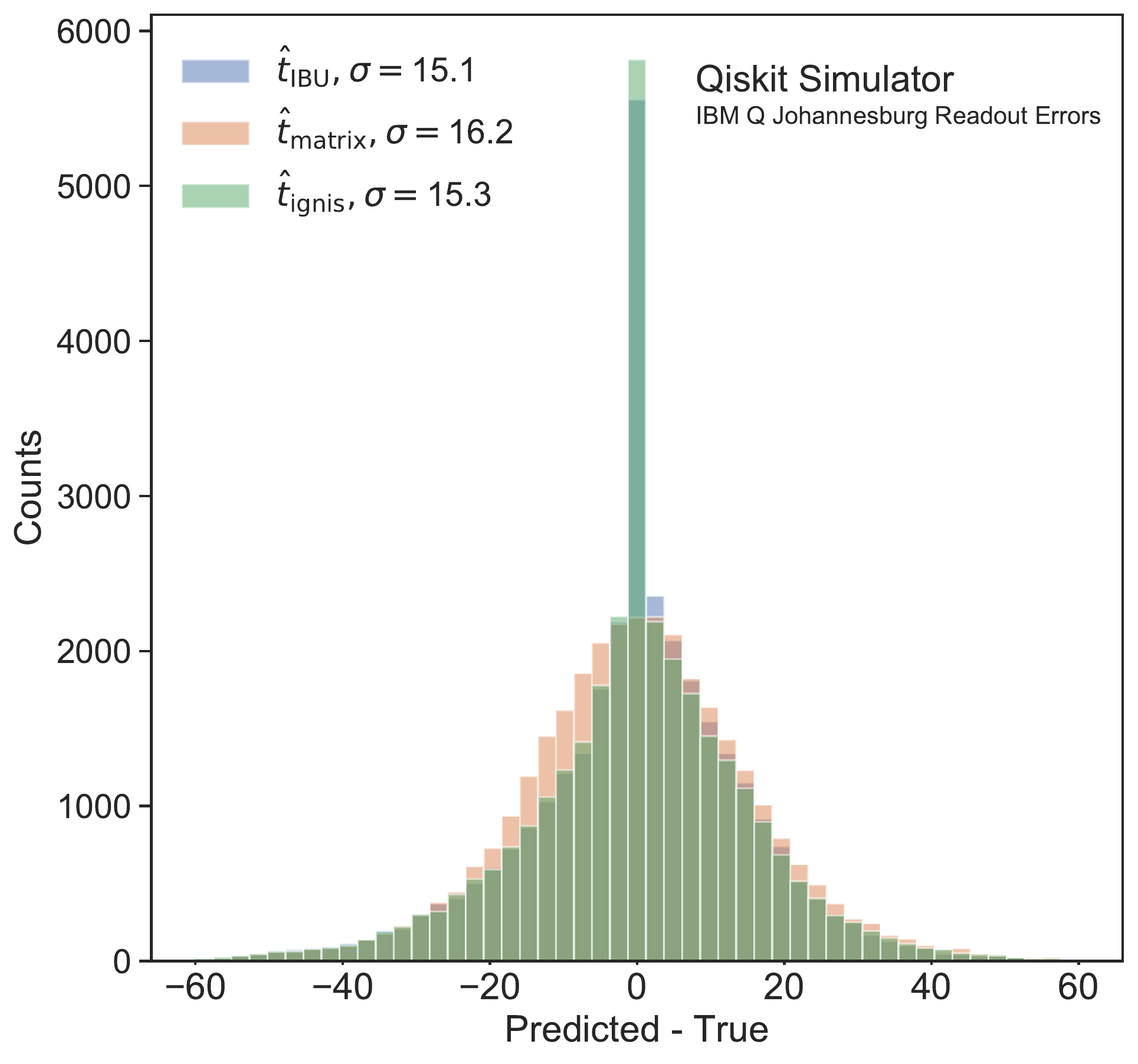}
\caption{The distribution of the difference between true and predicted counts from a Gaussian distribution using the response matrix from the IBM Q Johannesburg machine described in Sec.~\ref{sec:responsematrix}.  The simulation is repeated 1000 times.  For each of the 1000 pseudo-experiments, a discretized Gaussian state is prepared with no gate errors and it is measured $10^4$ times.  The bin counts without readout errors are the `true' values.  The measured distribution with readout errors is unfolded and the resulting counts are the `predicted' values.  Each of the $2^5$ states over all 1000 pseudo-experiments contribute one entry to the above histogram. The standard deviations of the distributions are given in the legend.  The IBU method uses 100 iterations.}
\label{lab:pull}
\end{figure}

The three unfolding approaches are quantitatively compared in Fig.~\ref{lab:pull}.  The same setup as Fig.~\ref{lab:pathologicalexampleqiskit} is used, repeated over 1000 pseudo-experiments.  For each of the 1000 pseudo-experiments, a discretized Gaussian state is prepared (same as Fig.~\ref{lab:pathologicalexampleqiskit}) with no gate errors and it is measured $10^4$ times.  The bin counts without readout errors are the `true' values.  These values include the irreducible Poisson noise that the unfolding methods are not expected to correct.  The measured distribution with readout errors is unfolded using  the three method and the resulting counts are the `predicted' values.  Averaging over many pseudo-experiments, the spread in the predictions is a couple of percent smaller for $\hat{t}_\text{IBU}$ compared to $\hat{t}_\text{ignis}$ and both of these are about 10\% more precise than $\hat{t}_\text{matrix}$.  The slight bias of $\hat{t}_\text{ignis}$ and $\hat{t}_\text{IBU}$ to the right results from the fact that they are non-negative.  Similarly, the sharp peak at zero results from $\hat{t}_\text{ignis}$ and $\hat{t}_\text{IBU}$ values that are nearly zero when $t_i\sim 0$.  In contrast, zero is not special for matrix inversion so there is no particular feature in the center of its distribution.

\begin{comment}

\subsection{Pathological Response Matrix}

\begin{figure}
\centering
\includegraphics[width=\linewidth]{fig4.pdf}
\caption{To be filled in.}
\label{lab:pathologicalexample}
\end{figure}

\subsection{Realistic Response Matrix}

\begin{comment}
\begin{figure}
\centering
\includegraphics[width=\linewidth]{fig3.pdf}
\caption{To be filled in.}
\label{lab:pathologicalexampleresponse}
\end{figure}

\begin{figure}
\centering
\includegraphics[width=\linewidth]{fig5.pdf}
\caption{To be filled in.}
\label{lab:pathologicalexampleresponse_approx}
\end{figure}
\end{comment}

\section{Regularization and Uncertainties}
\label{sec:uncerts}

\begin{figure}
\centering
\includegraphics[width=\linewidth]{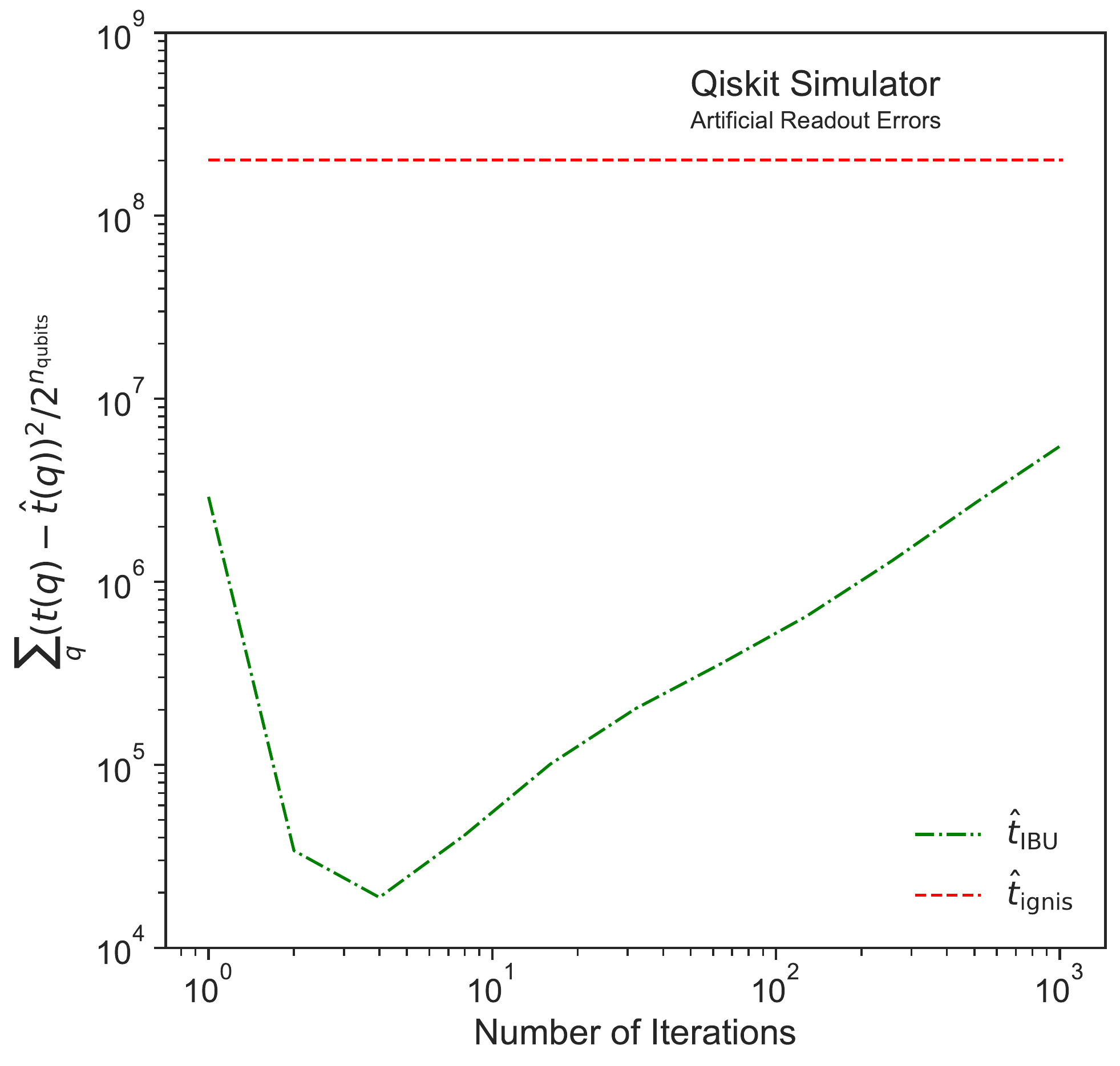}
\caption{The difference between $\hat{t}_\text{IBU}$ and $t$ as a function of the number of iterations for the example presented in Fig.~\ref{lab:pathologicalexampleqiskit}.  By definition, the \texttt{ignis} method does not depend on the number of iterations.}
\label{lab:uncerts0}
\end{figure}

\begin{figure}
\centering
\includegraphics[width=\linewidth]{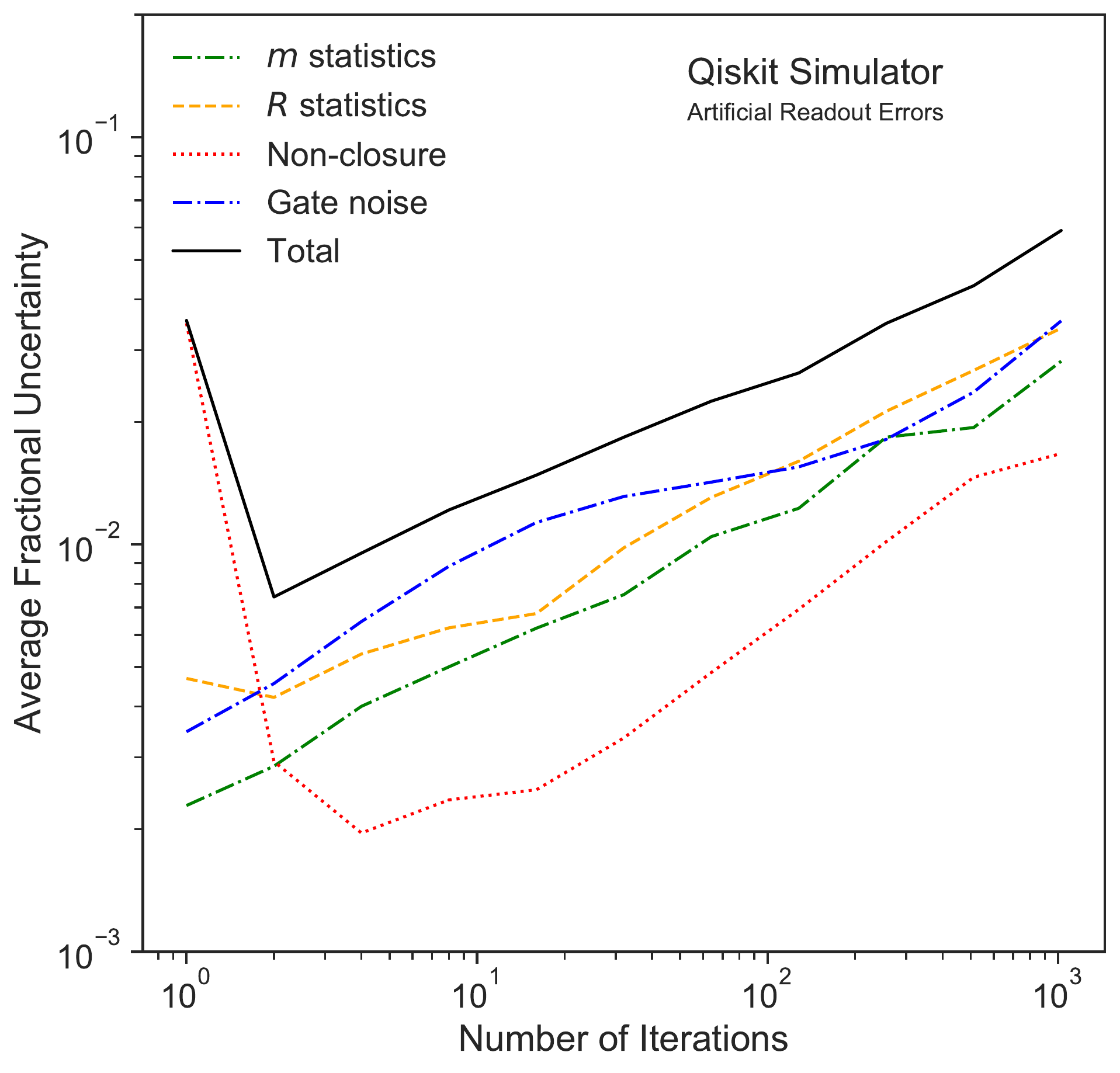}
\caption{Sources of uncertainty for $\hat{t}_\text{IBU}$ as a function of the number of iterations for the example presented in Fig.~\ref{lab:pathologicalexampleqiskit}.  Each uncertainty is averaged over all states.  The total uncertainty is the sum in quadrature of all the individual sources of uncertainty, except gate noise (which is not used in the measurement simulation, but would be present in practice).}
\label{lab:uncerts1}
\end{figure}

One feature of any regularization method is the choice of regularization parameters.  For the IBU method, these parameters are the prior and number of iterations.  Fig.~\ref{lab:uncerts0} shows the average bias from statistical and systematic sources in the measurement from the Gaussian example shown in Fig.~\ref{lab:pathologicalexampleqiskit} as a function of the number of iterations.  With a growing number of iterations, $\hat{t}_\text{IBU}$ approaches the oscillatory $\hat{t}_\text{ignis}$.  The optimal number of iterations from the point or view of the bias is 3.  However, the number of iterations cannot be chosen based on the actual bias because the true answer is not known a priori.  In HEP, the number of iterations is often chosen before \textit{unblinding} the data by minimizing the total expected uncertainty.  In general, there are three sources of uncertainty: statistical uncertainty on $m$, statistical and systematic uncertainties on $R$, and non-closure uncertainties from the unfolding method.  Formulae for the statistical uncertainty on $m$ are presented in Ref.~\cite{DAgostini:1994fjx} and can also be estimated by bootstrapping~\cite{efron1979} the measured counts.  Similarly, the statistical uncertainty on $R$ can be estimated by bootstrapping and then repeating the unfolding for each calibration pseudo-dataset.  The sources of statistical uncertainty are shown as dot-dashed and dashed lines in Fig.~\ref{lab:uncerts1}.  Adding more iterations enhances statistical fluctuations and so these sources of uncertainty increase monotonically with the number of iterations.

The systematic uncertainty on $R$ and the method non-closure uncertainty are not unique and require careful consideration.  In HEP applications\footnote{In HEP,  there are additional uncertainties related to the modeling of background processes as well as related to events that fall into or out of the measurement volume.  These are not relevant for QIS.}, $R$ is usually determined from simulation, so the systematic uncertainties are simulation variations that try to capture potential sources of mis-modeling.  These simulation variations are often estimated from auxiliary measurements with data.  In the QIS context, $R$ is determined directly from the data, so the only uncertainty is on the impurity of the calibration circuits.  In particular, the calibration circuits are constructed from a series of single qubit $X$ gates.  Due to gate imperfections and thermal noise, there is a chance that the application of an $X$ gate will have a different effect on the state than intended.  In principle, one can try to correct for such potential sources of bias by an extrapolation method.  In such methods, the noise is increased in a controlled fashion and then extrapolated to zero noise~\cite{Dumitrescu:2018,Kandala:2019,PhysRevX.7.021050,PhysRevLett.119.180509,PhysRevX.8.031027}.  This method may have a residual bias and the uncertainty on the method would then become the systematic uncertainty on $R$.  A likely conservative alternative to this approach is to modify $R$ by adding in gate noise and taking the difference between the nominal result and one with additional gate noise, simulated using the \texttt{thermal}\_\texttt{relaxation}\_\texttt{error} functionality of \texttt{qiskit}.  This is the choice made in Fig.~\ref{lab:uncerts1}, where the gate noise is shown as a dot-dashed line.  In this particular example, the systematic uncertainty on $R$ increases monotonically with the number of iterations, just like the sources of statistical uncertainty.

The non-closure uncertainty is used to estimate the potential bias from the unfolding method.  One possibility is to compare multiple unfolding methods and take the spread in predictions as an uncertainty.  Another method advocated in Ref.~\cite{Malaescu:2009dm} and widely used in HEP is to perform a data-driven reweighting.  The idea is to reweight the $t^0$ so that when folded with $R$, the induced $m^0$ is close to the measurement $m$.  Then, this reweighted $m^0$ is unfolded with the nominal response matrix and compared with the reweighted $t^0$.  The difference between these two is an estimate of the non-closure uncertainty.  The reweighting function is not unique, but should be chosen so that the reweighted $t^0$ is a reasonable prior for the data.  For Fig.~\ref{lab:uncerts1}, the reweighting is performed using the nominal unfolded result itself.  In practice, this can be performed in a way that is blinded from the actual values of $\hat{t}_\text{IBU}$ so that the experimenter is not biased when choosing the number of iterations. 

Altogether, the sources of uncertainty presented in Fig.~\ref{lab:uncerts1} show that the optimal choice for the number of iterations is 2.  In fact, the difference in the uncertainty between 2 and 3 iterations is less than 1\% and so consistent with the results from Fig.~\ref{lab:uncerts0}.  Similar plots for the measurement in Fig.~\ref{lab:nonpathologicalexampleqiskit} can be found in Appendix~\ref{sec:uncerts2}.

\section{Discussion}
\label{sec:discussion}

This work has introduced a new suite of readout error correction algorithms developed in high energy physics for binned differential cross section measurements.  These unfolding techniques are well-suited for quantum computer readout errors, which are naturally binned and without acceptance effects (counts are not lost or gained during readout).  In particular, the iterative Bayesian method has been described in detail and shown to be robust to a failure mode of the matrix inversion and \texttt{ignis} techniques.  When readout errors are sufficiently small, all the methods perform well, with a preference for the \texttt{ignis} and Bayesian methods that produce non-negative results.  The \texttt{ignis} method is a special case of the \texttt{TUnfold} algorithm, where the latter uses the covariance matrix to improve precision and incorporates regularization to be robust to the failure modes of matrix inversion.  It may be desirable to augment the \texttt{ignis} method with these features or provide the iterative method as an alternative approach.  In either case, Fig.~\ref{lab:nonpathologicalexampleqiskit} showed that even with a realistic response matrix, readout error corrections can be significant and must be accounted for in any measurement on near-term hardware.

An important challenge facing any readout error correction method is the exponential resources required to construct the full $R$ matrix as mentioned in Sec.~\ref{sec:responsematrix}.  While $R$ must be constructed only once per hardware setup and operating condition, it could become prohibitive when the number of qubits is large.  On hardware with few connections between qubits, per-qubit transition probabilities may be sufficient for accurate results.  When that is not the case, one may be able to achieve the desired precision with polynomially many measurements.  These ideas are left to future studies.

Another challenge is optimizing the unfolding regularization.  Section~\ref{sec:uncerts} considered various sources of uncertainty and studied how they depend on the number of iterations in the IBU method.  The full measurement is performed for all $2^{n_\text{qubit}}$ states and the studies in Sec.~\ref{sec:uncerts} collapsed the uncertainty across all bins into a single number by averaging across bins.  This way of choosing a regularization parameter is common in HEP, but is not a unique approach.  Ultimately, a single number is required for optimization, but it may be that other metrics are more important for specific applications, such as the uncertainty for a particular expectation value or the maximum or most probable uncertainty across states.  Such requirements are application specific, but should be carefully considered prior to the measurement.

\section{Conclusions and Outlook}
\label{sec:conclusions}

With active research and development across a variety of application domains, there are many promising applications of quantum algorithms in both science and industry on NISQ hardware.  Readout errors are an important source of noise that can be corrected to improve measurement fidelity.  High energy physics experimentalists have been studying readout error correction techniques for many years under the term unfolding.  These tools are now available to the QIS community and will render the correction procedure more robust to resolution effects in order to enable near term breakthroughs.

\begin{acknowledgments}

This work is supported by the U.S. Department of Energy, Office of Science under contract DE-AC02-05CH11231. In particular, support comes from Quantum Information Science Enabled Discovery (QuantISED) for High Energy Physics (KA2401032) and the Office of Advanced Scientific Computing Research (ASCR) through the Quantum Algorithms Team program. We acknowledge access to quantum chips and simulators through the IBM Quantum Experience and Q Hub Network through resources of the Oak Ridge Leadership Computing Facility, which is a DOE Office of Science User Facility supported under Contract DE-AC05-00OR22725.  BN would also like to thank Michael Geller for spotting a typo, Jesse Thaler for stimulating discussions about unfolding, and the Aspen Center for Physics, which is supported by National Science Foundation grant PHY-1607611.

\end{acknowledgments}

\section*{Author Contributions}

BN conceived the project idea, wrote the code, performed the numerical analysis, and wrote the manuscript.  WD organized the measurements on IBMQ.  All authors discussed the results and revised the manuscript. 

\section*{Competing Interests}

The authors declare that there are no competing interests.

\section*{Data Availability}

The code for the work presented here can be found at \texttt{https://github.com/bnachman/QISUnfolding}.

\appendix

\section{Iteration-Dependence of IBU}
\label{sec:iterations}

\begin{figure*}
\centering
\includegraphics[width=0.5\linewidth]{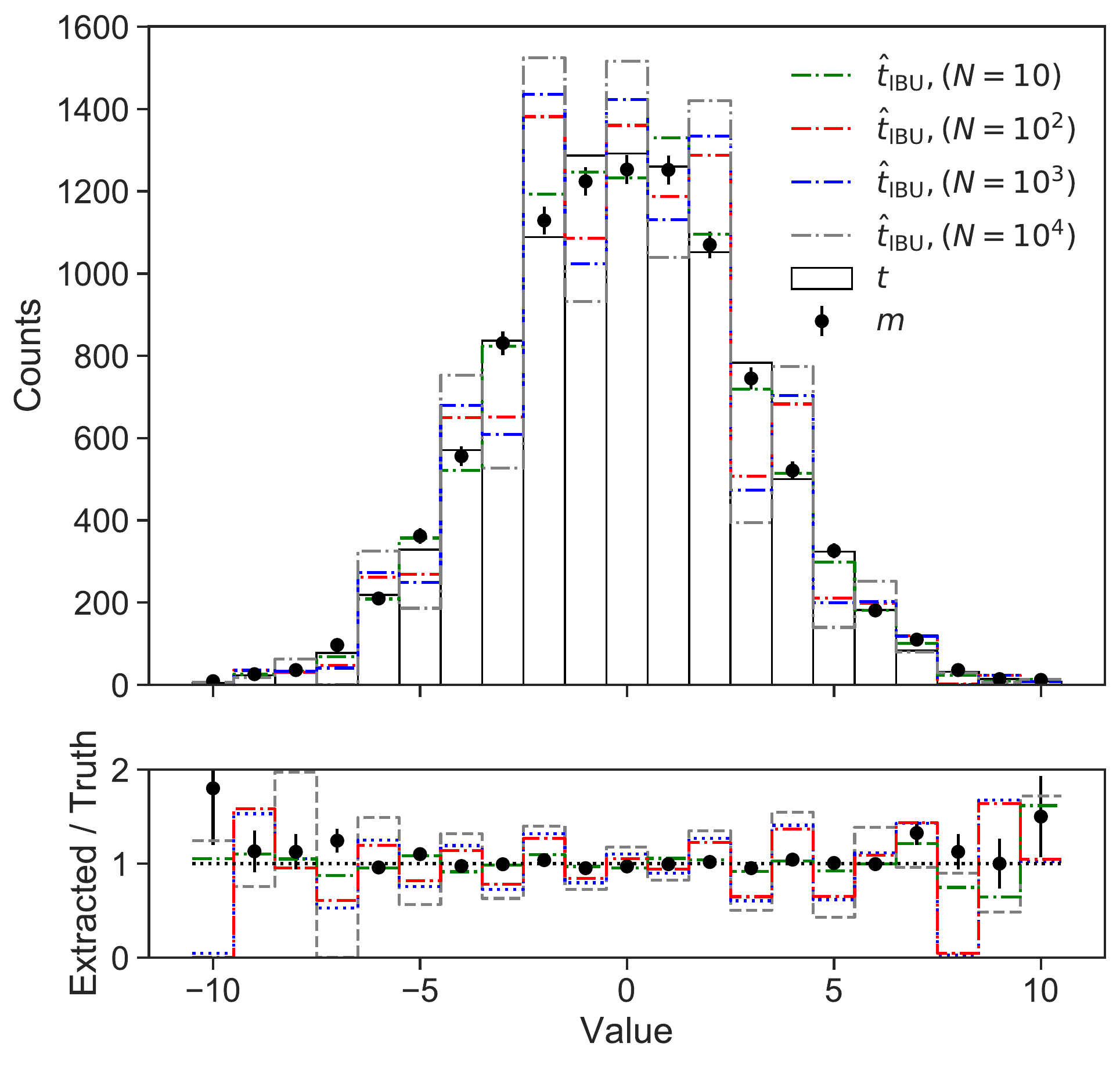}
\caption{The same as Fig.~\ref{lab:matrixinversion}, but with increasingly many iterations ($N$) for IBU.}
\label{lab:iterations}
\end{figure*}

Figure~\ref{lab:iterations} shows the impact of adding more iterations for IBU to the plot that appeared in Fig.~\ref{lab:matrixinversion}.  As the number of iterations increases, the oscillation behavior of the other algorithms starts to appear.  Stopping the iterative procedure early is a regularization that can help damp oscillations that are known to form when $N\rightarrow\infty$.

\section{Alternative Configurations of the IBM Q Johannesburg Machine}
\label{sec:alternativeconfigurations}

\begin{figure*}
\centering
\includegraphics[width=0.5\linewidth]{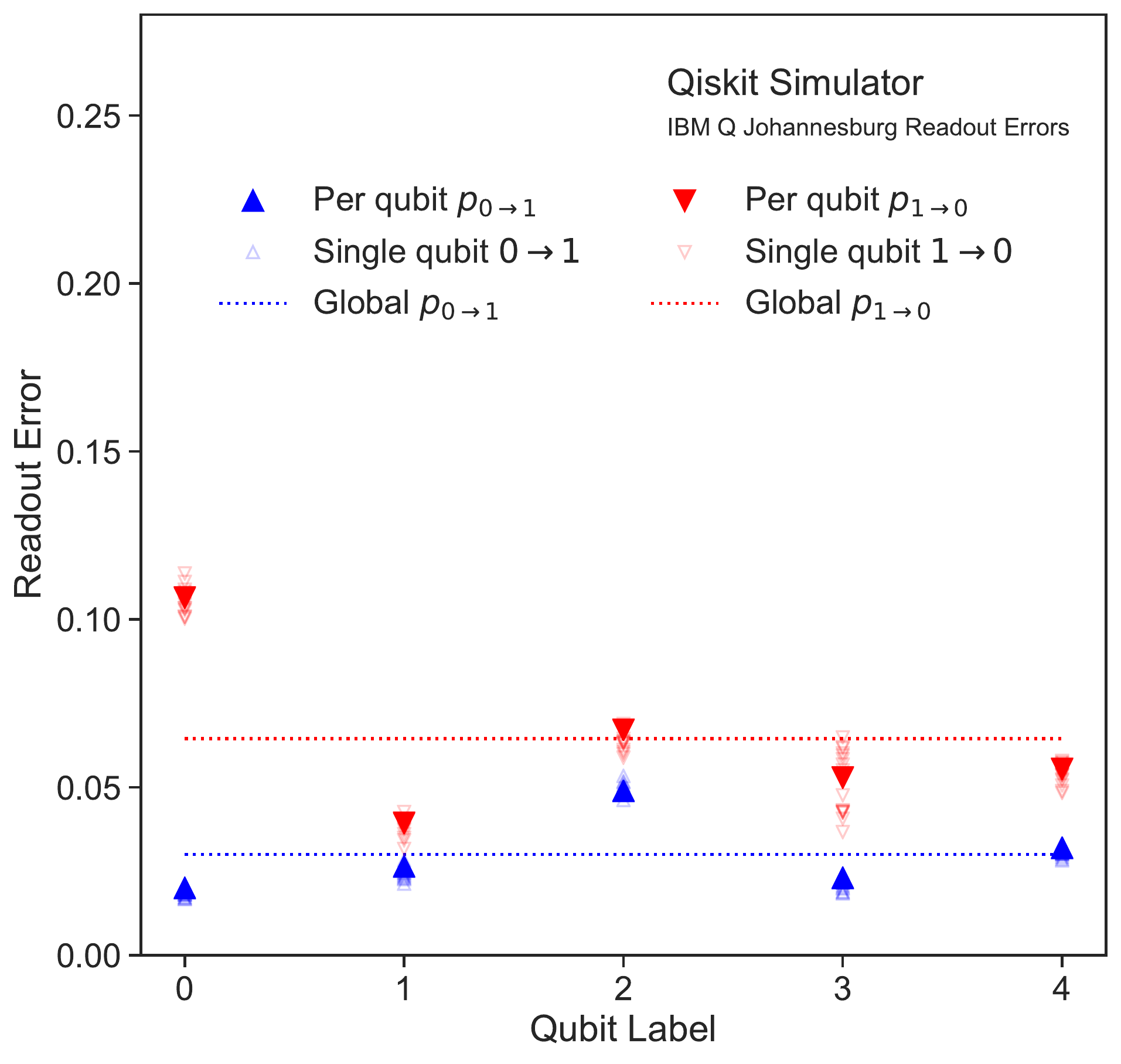}\includegraphics[width=0.5\linewidth]{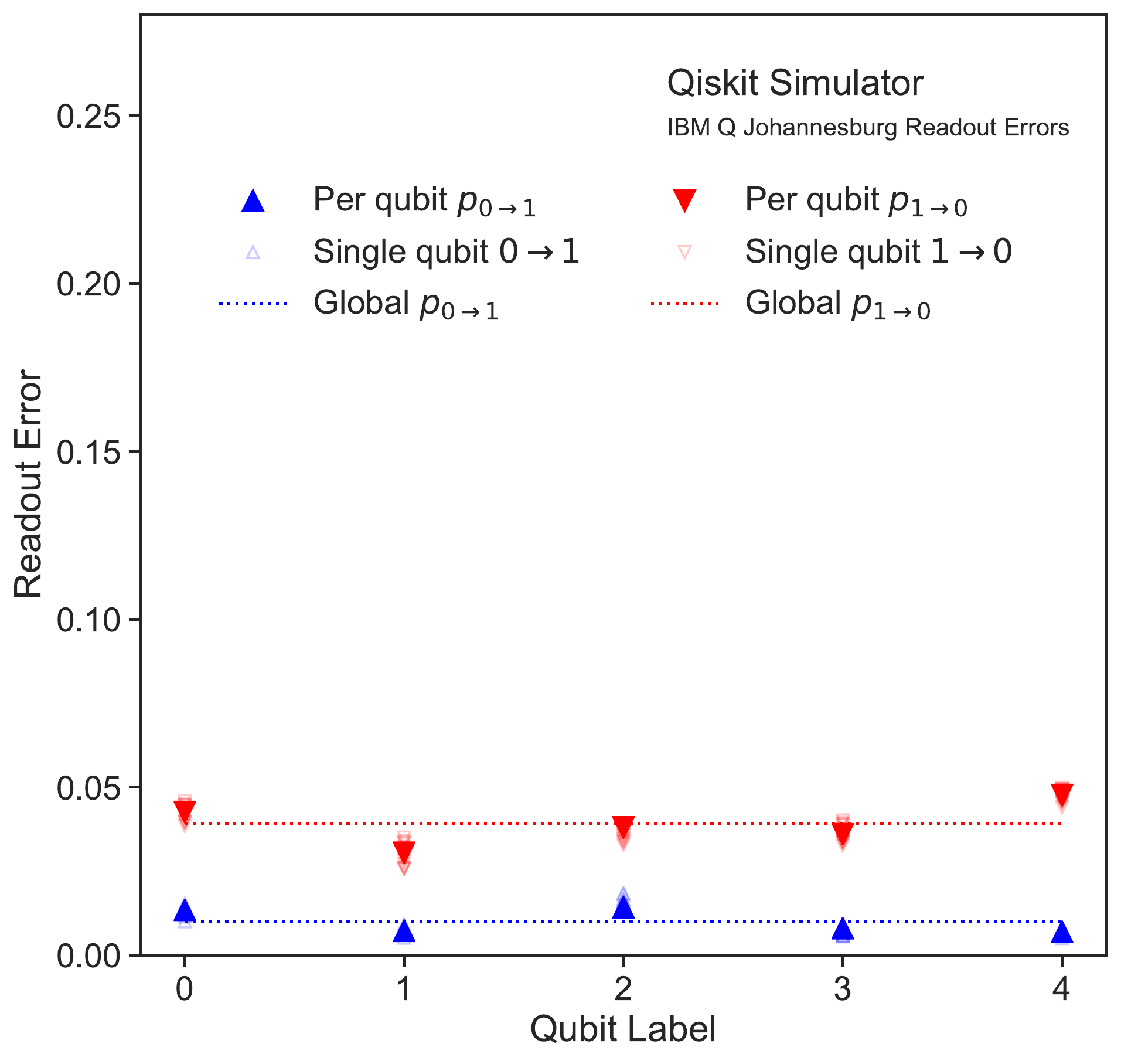}\\
\includegraphics[width=0.5\linewidth]{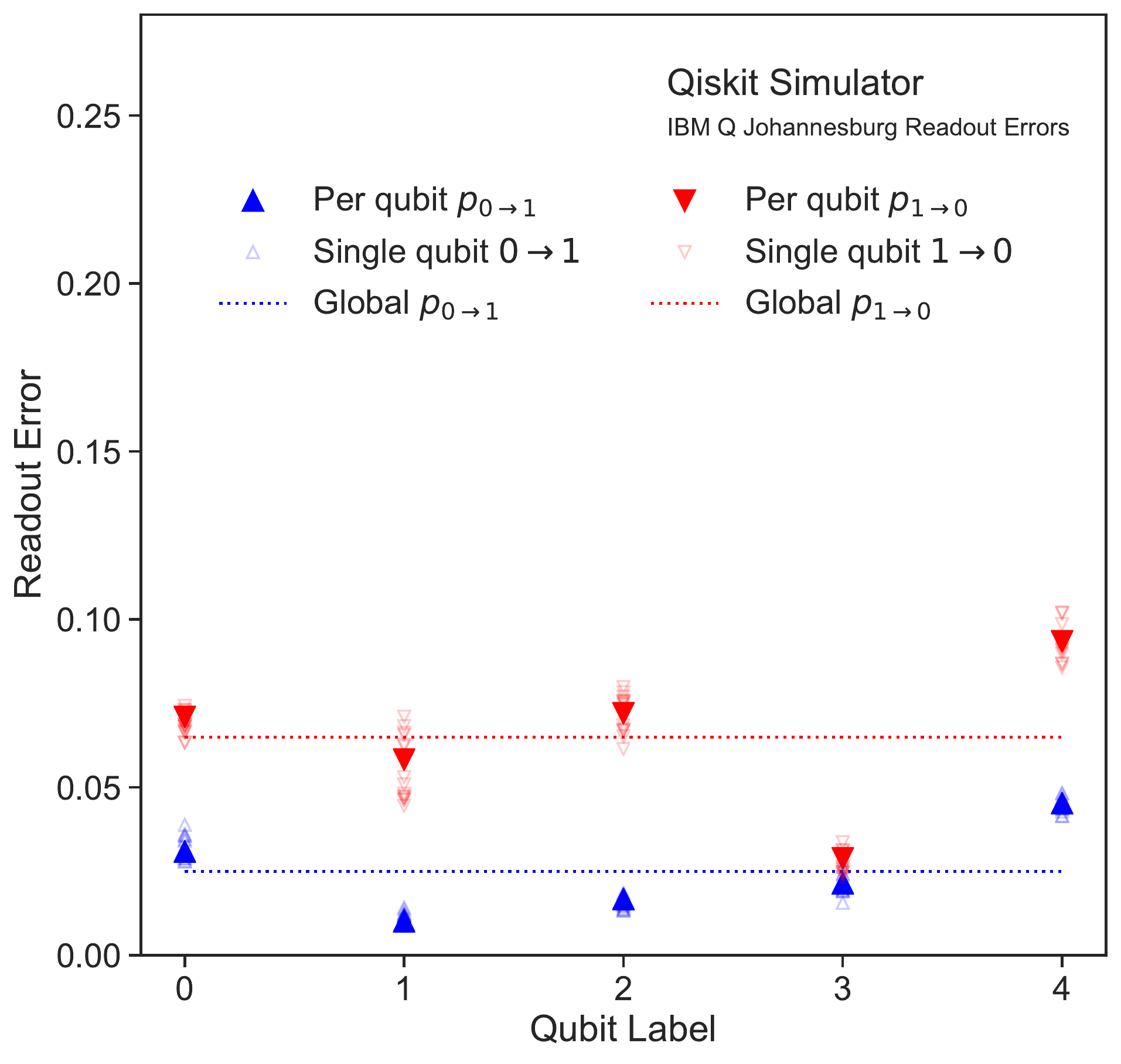}\includegraphics[width=0.5\linewidth]{fig5_4.pdf}
\caption{The same test of universality as in Fig.~\ref{lab:universality}, using various configurations of the IBM Q Johannesburg machine.  From top left, clockwise, the plots correspond to rows 1-4 as described in the text.}
\label{lab:universality2}
\end{figure*}

The readout errors of a quantum computer depend on its connectivity.  The results presented in Fig.~\ref{lab:universality} used a particular subset of 5 qubits out of the 20 qubits available.  Fig.~\ref{lab:universality2} presents the results for four different configurations, corresponding to taking four rows of the computer qubits.  In each row, every qubit is connected to its two neighbors.  The first and last qubit in each row is connected to the row above or below.  The middle qubits in the middle two rows are additionally connected to each other.  Therefore, every qubit in the first and last rows have only two connections while two of the qubits in the middle rows have two connections and the other three qubits have three connections.

\section{Uncertainties when using IBM Q Response Matrix}
\label{sec:uncerts2}

\begin{figure}
\centering
\includegraphics[width=\linewidth]{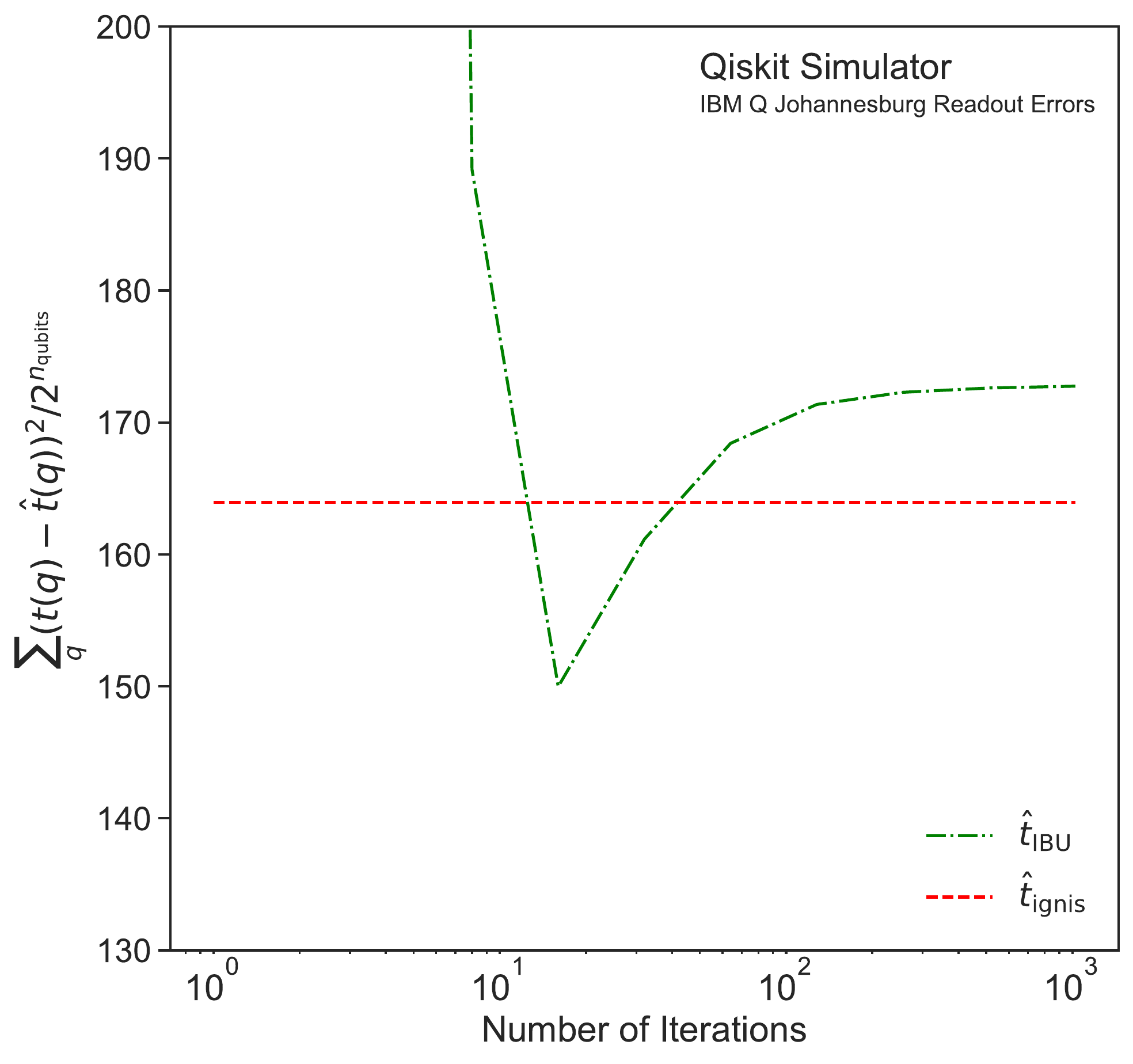}
\caption{The difference between $\hat{t}_\text{IBU}$ and $t$ as a function of the number of iterations for the example presented in Fig.~\ref{lab:nonpathologicalexampleqiskit}.  By definition, the \texttt{ignis} method does not depend on the number of iterations.}
\label{lab:uncerts0r}
\end{figure}

\begin{figure}
\centering
\includegraphics[width=\linewidth]{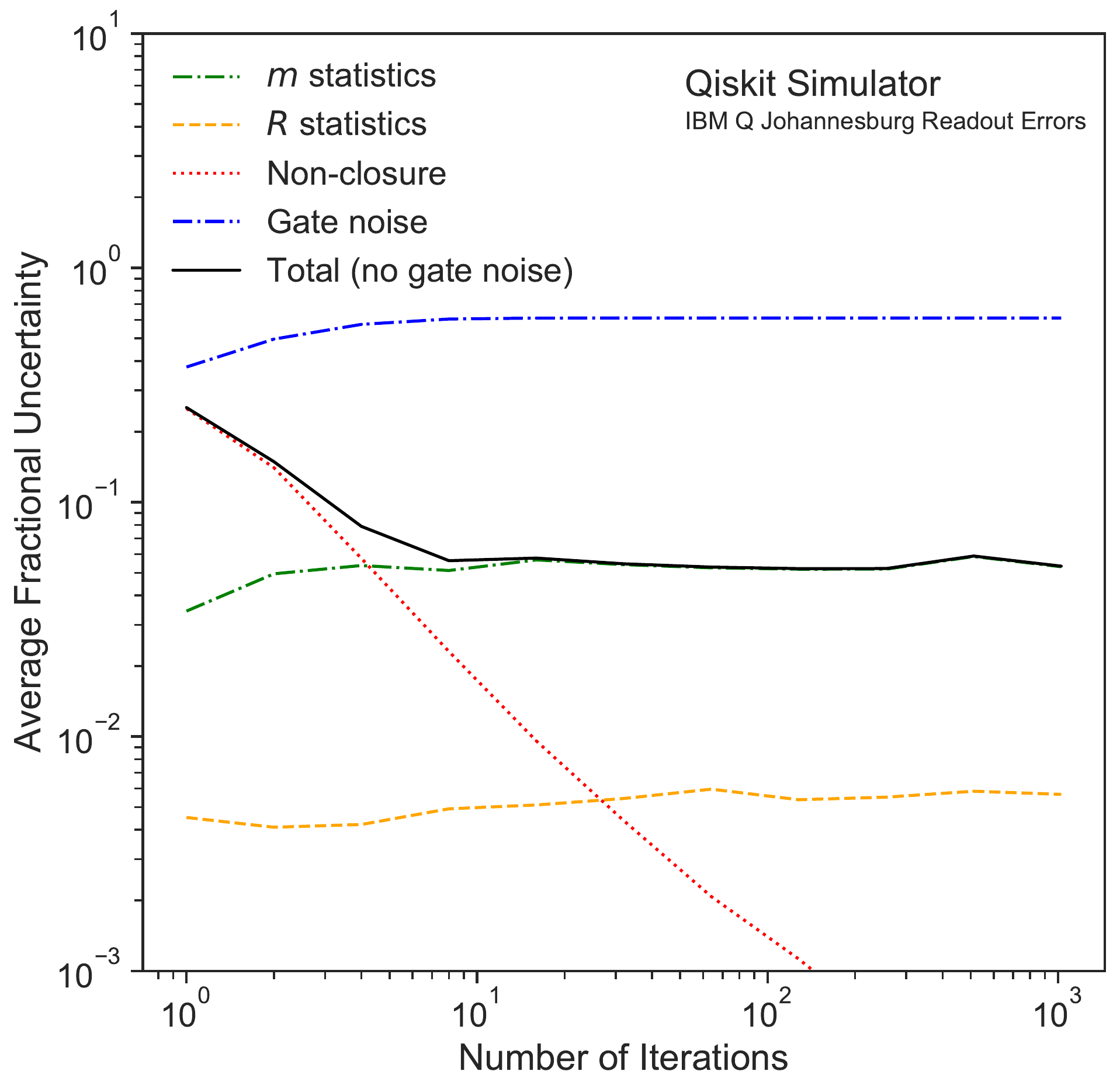}
\caption{Sources of uncertainty for $\hat{t}_\text{IBU}$ as a function of the number of iterations for the example presented in Fig.~\ref{lab:nonpathologicalexampleqiskit}.  Each uncertainty is averaged over all states.  The total uncertainty is the sum in quadrature of all the individual sources of uncertainty, except gate noise (which is not used in the measurement simulation, but would be present in practice).}
\label{lab:uncerts1r}
\end{figure}

Fig.~\ref{lab:uncerts0r} and~\ref{lab:uncerts1r} are the analogs for Fig.~\ref{lab:uncerts0} and~\ref{lab:uncerts1}, but for the example shown in Fig.~\ref{lab:nonpathologicalexampleqiskit}, which is based on an IBM Q response matrix.

\section{Alternative to Harmonic Oscillator Ground State}
\label{sec:alternative}

\begin{figure}
\centering
\includegraphics[width=\linewidth]{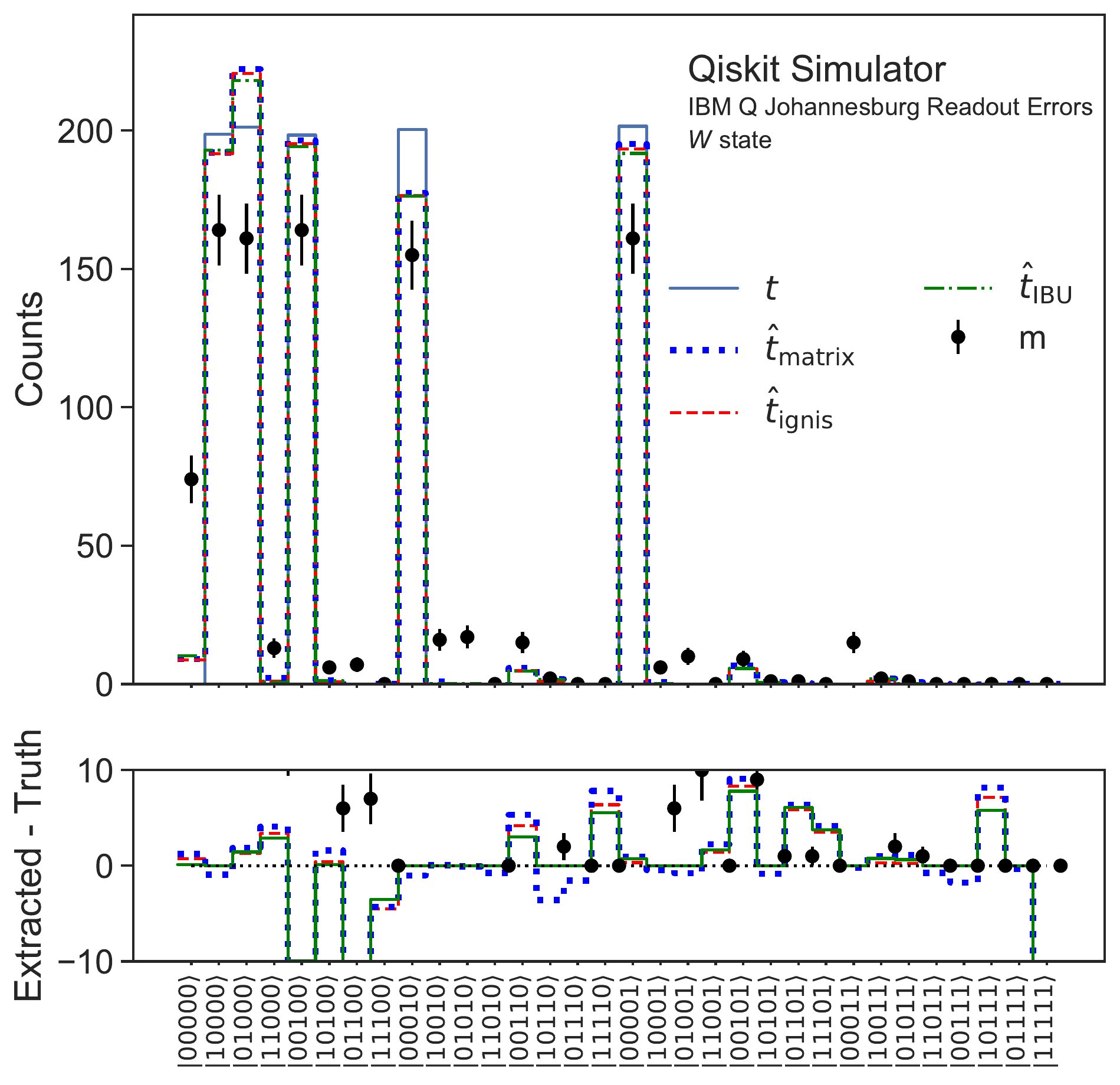}
\caption{The measurement of the noisy $W$ distribution using the response matrix from the IBM Q Johannesburg machine described in Sec.~\ref{sec:responsematrix}.  1000 shots are used for the state measurement and $500\times 2^{n_\text{qubits}}$ shots are used to construct the response matrix.  IBU uses 100 iterations.  Note that unlike for Fig.~\ref{lab:pathologicalexampleqiskit} and~\ref{lab:nonpathologicalexampleqiskit}, the true distribution is the theoretical $W$ distribution and not the true pre-readout-noise distribution.   This is because the $W$ state is prepared as fully entangled prior to measurement.}
\label{lab:nonpathologicalexampleqiskit2}
\end{figure}

Fig.~\ref{lab:nonpathologicalexampleqiskit2} presents a spiky distribution instead of a smooth probability density, as typified by the ground state of the harmonic oscillator.   The state corresponds to the $W$ distribution~\cite{PhysRevA.62.062314}: $\ket{W} = \frac{1}{\sqrt{n}}\left(\ket{100\cdots 0}+\ket{010\cdots 0}+\cdots\ket{00\cdots 01}\right)$.  As the true distribution is exactly zero for most states, the matrix inversion has large fluctuations because there is nothing special about zero for this method (can give negative results).  In contrast, the \texttt{ignis} and IBU methods are always positive and have smaller deviations.  Figure~\ref{nonpathologicalexampleqiskit2_bias} shows that the mean squared error for IBU is lower than \texttt{ignis} for tens of iterations.

%For this example, the IBU method p.  This also means that the number of iterations required to minimize the bias is much larger than the Gaussian case - see Fig.~\ref{lab:uncerts0r2}.

\begin{figure}
\centering
\includegraphics[width=\linewidth]{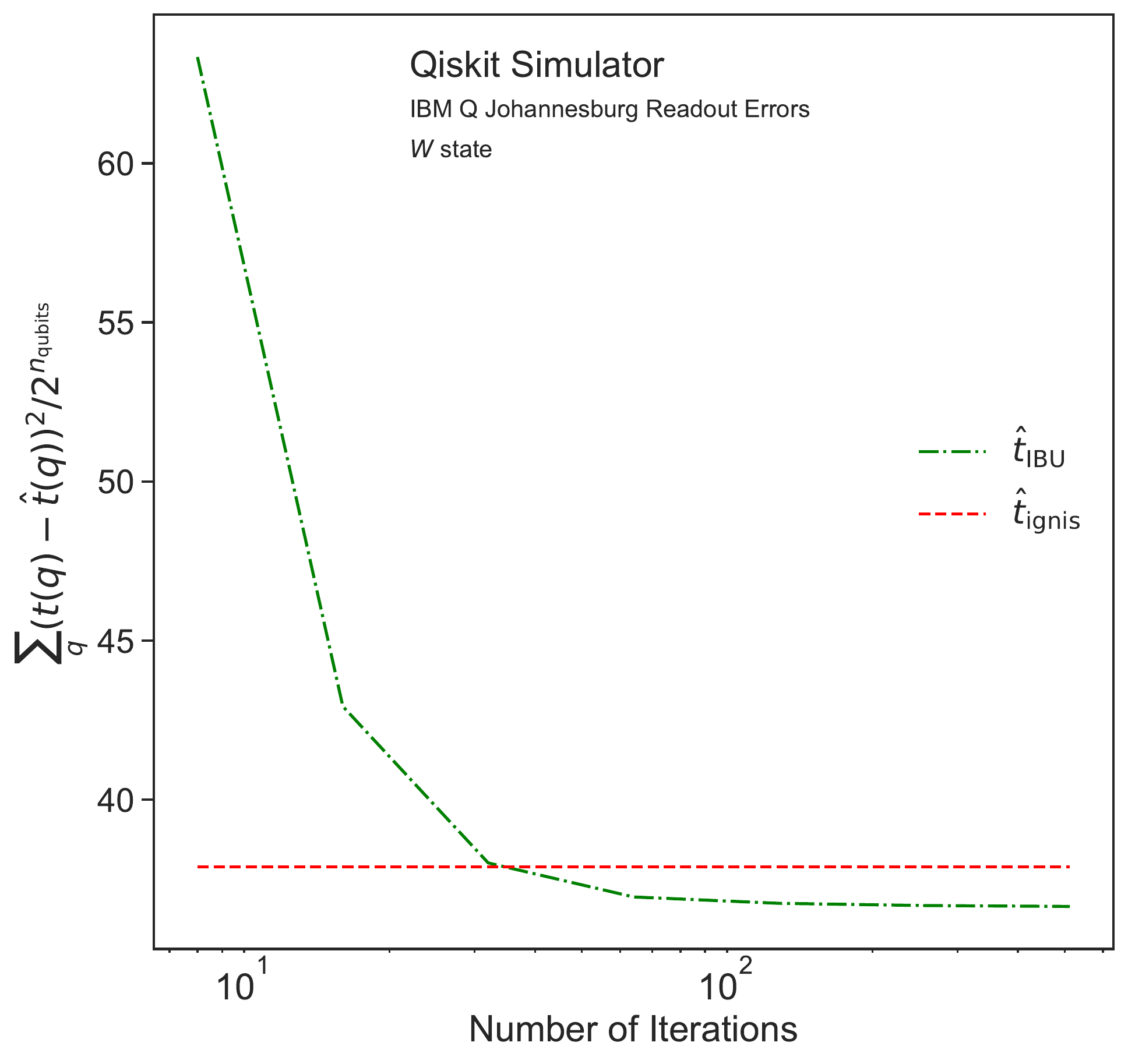}
\caption{The mean squared error between the unfolded and theoretical $W$ state distribution as a function of the number of iterations in the IBU method.}
\label{lab:nonpathologicalexampleqiskit2_bias}
\end{figure}

%\begin{figure}
%\centering
%\includegraphics[width=\linewidth]{newfig_bias_r2.pdf}
%\caption{The absolute value of the average difference between the estimate $\hat{t}$ and $t$ as a function of the number of iterations in the IBU method for the example presented in Fig.~\ref{lab:nonpathologicalexampleqiskit2}.  By definition, the \texttt{ignis} method does not depend on the number of iterations.}
%\label{lab:uncerts0r2}
%\end{figure}

\clearpage

\bibliographystyle{apsrev4-1}
\bibliography{myrefs}

\end{document}